# A HR-like Diagram for Solar/Stellar Flares and Corona – Emission Measure vs Temperature Diagram

Kazunari Shibata[1] and Takaaki Yokoyama[2]

[1]: Kwasan and Hida Observatories, Kyoto University, Yamashina, Kyoto 607-8471, Japan, shibata@kwasan.kyoto-u.ac.jp

[2] National Astronomical Observatory, Nobeyama, Minamimaki, Minamisaku, Nagano 384-1305, Japan, yokoyama.t@nao.ac.jp

## ABSTRACT

In our previous paper, we have presented a theory to explain the observed universal correlation between the emission measure ($EM = n^2 V$) and temperature ($T$) for solar/stellar flares on the basis of the magnetic reconnection model with heat conduction and chromospheric evaporation. Here $n$ is the electron density and $V$ is the volume. By extending our theory to general situations, we examined the $EM$-$T$ diagram in detail, and found the following properties: 1) The universal correlation sequence ("main sequence flares") with $EM \propto T^{17/2}$ corresponds to the case of constant heating flux or equivalently the case of constant magnetic field strength in the reconnection model. 2) The $EM$-$T$ diagram has a forbidden region, where gas pressure of flares exceeds magnetic pressure. 3) There is a coronal branch with $EM \propto T^{15/2}$ for $T < 10^7$ K, and $EM \propto T^{13/2}$ for $T > 10^7$ K. This branch is situated left side of the main sequence flares in the $EM$-$T$ diagram. 4) There is another forbidden region determined by the length of flare loop; a lower limit of flare loop is $10^7$ cm. Small flares near this limit correspond to nanoflares observed by SOHO/EIT. 5) We can plot flare evolution track on the $EM$-$T$ diagram. A flare evolves from the coronal branch to main sequence flares, then returns to the coronal branch eventually. These properties of the $EM$-$T$ diagram are similar to those of the HR diagram for stars, and thus we propose that the $EM$-$T$ diagram is quite useful to estimate the physical quantities (loop length, heating flux, magnetic field strength, total energy and so on) of flares and corona when there is no spatially resolved imaging observations.

*Subject headings:* stars:flares – X-rays:stars – MHD

## 1. Introduction

Recent space observations of solar and stellar flares have revealed that there is a universal correlation between the peak temperature ($T$) of a flare and its volume emission measure ($EM = n^2 V$), such that $EM$ increases with increasing $T$, where $n$ is the electron number density and $V$ is the volume (Watanabe 1983, Stern 1992, Feldman et al. 1995, 1996, Yuda et al. 1997). This correlation is extrapolated to not only solar microflares (Shimizu 1995) but also protostellar flares (Koyama et al. 1996, Tsuboi et al. 1998, Tsuboi 1999). Figure 1 shows the $EM$-$T$ relation for solar and stellar flares, including solar microflares and protostellar flares. It is remarkable that the correlation holds for wide range of parameter values $10^{44} < EM < 10^{56}$ cm$^{-3}$ and $6 \times 10^6 < T < 10^8$ K.

Recently, Shibata and Yokoyama (1999) presented a simple theory to explain this remarkable correlation on the basis of the flare temperature scaling law, which was found from magnetohydrodynamic (MHD) numerical simulations of reconnection coupled with heat conduction and chromospheric evaporation (Yokoyama and Shibata 1998, 2001). Shibata and Yokoyama (1999) derived theoretical correlation, $EM \propto B^{-5} T^{17/2}$, which explains observed correlation if the magnetic field strength $B$ is nearly constant within a factor of 10 for solar and stellar flares (Fig. 2). In fact, magnetic field strengths of solar and stellar flares are estimated to be 40-300 G (e.g., Rust and Bar 1973, Dulk 1985, Tsuneta 1996, Ohyama and Shibata 1998, Grosso et al. 1997), consistent with above theory. This means that the observed $EM$-$T$ diagram may be useful to estimate magnetic field strength of stellar flares, which are difficult to observe.

Shibata and Yokoyama (1999) noticed also that this theory can be used to estimate the flare loop length. In fact, from the comparison between the theory and the observed $EM$-$T$ relation, they estimated that the loop length of solar flares is $10^9 - 10^{10}$ cm, and that of stellar flares is $10^{11} - 10^{12}$ cm, which are again consistent with direct observations of solar flare loops and with some indirect estimate of stellar flare loops (Koyama et al. 1996, Hayashi, Shibata, Matsumoto 1996, Montmerle 1998; see also Haisch et al. 1991, Feigelson and Montmerle 1999 for a review of stellar flare observations).

As for estimate of the stellar flare loop length, Reale and Micera (1998) and, more recently, Favata, Micela, and Reale (2001) took into account both the flare impulsive phase and its decay, allowing for continuous energy deposition, and derived the size of the flare loop by calibrating the slope of the light curve in the $EM$-$T$ diagram based on the hydrodynamic modeling (Serio et al. 1991). They took also into account the pass band of the instrument used to derive the $EM$, and found smaller loop sizes than the $\sim 10^{11}$ cm obtained in previous papers. Note that the method of estimating the flare loop used by previous papers (e.g., Montmerle 1998, Tsuboi et al. 2000) is based on the assumption that the flare decay can be described through a series of static states (i.e., the quasi-static model; van den Oord and Mewe 1989). On the other hand, our method (Shibata and Yokoyama 1999, and this paper) is based only on the peak flare temperature, the peak emission measure, and the preflare coronal density, so that it does not depend on any physical process occurring in the flare decay phase. We shall discuss this point in more detail in this paper.

It should be noted that the emission measure and temperature may be a bit ill defined quantities in a non-isothermal plasma such as in a flaring region. Their determination depends strongly on the pass pand of the instrument used for the observations (Reale and Micera 1998). We note that when we compare our theory and the observed data in this paper, we will not take into account such pass band effect. Hence too detailed comparison between the theory and the observed data such as within an accuracy of a factor of 2 or 3 may be meaningless, and the correlation between $EM$ and $T$ would be only statistically meaningful. We leave the more accurate comparison between the theory and the observations in future studies, which would incorporate the calibration of the pass band of the instrument used for observations. Nevertheless, in this paper, we emphasize the importance of the $EM$-$T$ correlation, because the $EM$-$T$ diagram for solar and stellar flares is in some sense similar to the HR diagram for stars. In the case of the HR diagram, if we plot a star on the HR diagram, we can infer the stellar radius without direct measurement. We can also discuss several branches of stars, such as main sequence, giants, white dwarfs, etc., and also their evolution in the HR diagram. Similarly, we can discuss several branches of flares and flare-like phnenomena and their evolution in the $EM$-$T$ diagram.

In this paper, noting the analogy between the flare $EM$-$T$ diagram and the HR diagram, we will examine basic physical properties of the $EM$-$T$ diagram for flares. In section 2, we summarize the derivation of the $EM$-$T$ correlation ($EM \propto B^{-5} T^{17/2}$) by Shibata and Yokoyama (1999) which was based on the reconnection model and the pressure balance of flare loops. We relax this pressure balance assumption and obtain a bit different relation ($EM \propto T^{15/2}$) in the case of enthalpy flux - conduction flux balance. The forbidden region is found in the $EM$-$T$ diagram, which reminds us of Hayashi's forbidden region in the HR diagram. In section 3, we investigate coronal branch in the $EM$-$T$ diagram where the heating is arbitrary and the radiative cooling becomes comparable to conduction flux, which predicts $EM \propto T^{15/2}$ for solar corona, and $EM \propto T^{13/2}$ for stellar corona with temperature higher than $10^7$ K. The coronal branch lies left side of the flare branch in the $EM$-$T$ diagram. In section 4, we discuss the flare evolution track on this $EM$-$T$ diagram, nanoflare branch, and the estimate of total flare energy using this diagram. Finally, section 5 is devoted to the conclusions.

We emphasize again that the *EM-T* diagram is useful to estimate currently-unobservable physical quantities such as magnetic field strength, flare loop length, conduction and radiative cooling times, etc. using only three observed quantities, peak emission measure $EM$, peak temperature $T$, and pre-flare coronal density $n_0$. The detailed time evolution data of stellar flares are not necessary in our method in contrast with other method such as the one based on the flare decay time (e.g., van den Oord and Mewe 1989, Reale and Micera 1998). The useful formula to estimate various physical quantities using $EM, T, n_0$ are summarized in Appendix 1. Limitation of the temperature scaling law (Yokoyama and Shibata 1998) and the condition of chromospheric evaporation are discussed in Appendix 2 and 3, respectively. The effect of the filling factor on the flare *EM-T* scaling law will be discussed in Appendix 4.

## 2. Flare Branch: Main Sequence

### 2.1. The Pressure Balance Scaling Law

The theory of Shibata and Yokoyama (1999) is based on a simple scaling relation for the maximum temperature of reconnection heated plasma $T_{max}$ and magnetic field strength $B$, which is found from MHD simulations by Yokoyama and Shibata (1998, 2001);

$$T_{max} \simeq \Big(\frac{B^2 v_A L}{\kappa_0 2\pi}\Big)^{2/7} \simeq 5.3 \times 10^4 \ B^{6/7} n_0^{-1/7} L^{2/7} \quad \text{K}$$

$$\simeq 3 \times 10^7 \Big(\frac{B}{50\text{G}}\Big)^{6/7} \Big(\frac{n_0}{10^9 \text{cm}^{-3}}\Big)^{-1/7} \Big(\frac{L}{10^9 \text{cm}}\Big)^{2/7} \quad \text{K}, \tag{1}$$

where $v_A = B/(4\pi\rho)^{1/2}$ is the Alfven speed, $\rho(= mn_0)$ is the mass density, $m$ is the proton mass, $n_0$ is the pre-flare proton number density (= electron density), $L$ is the characteristic length of the (reconnected) magnetic loop. This relation is derived from the balance between conduction cooling $\kappa_0 T^{7/2}/(2L^2)$ (e.g., Fisher and Hawley 1990, Hori et al. 1997) and reconnection heating $(B^2/4\pi)(v_A/L)$. Here, $\kappa_0 \simeq 10^{-6}$ CGS is the Spitzer (1956)'s thermal conductivity. In this model (see Figure 2 of Shibata and Yokoyama, 1999), after the reconnection occurs, the released energy is transported to the top of the chromosphere by heat conduction, and heats the chromospheric plasma suddenly. Then its plasma pressure increases enormously, and drives the upward flow back into the (reconnected) coronal magnetic loop, creating hot dense flare loops. (This process is often called "chromospheric evaporation".) As a result of the chromospheric evaporation, the flare loop density increases to $n$. This evaporated plasma with density $n$ is the source of X-ray emission;

$$EM \simeq n^2 L^3. \tag{2}$$

Here, we assumed $V \simeq L^3$ and filling factor $= 1$ (see Appendix 4 for the effect of filling factor), since Yohkoh/SXT observations of microflares and flares (Shimizu 1995, Takahashi 1997) show that observed aspect ratio of flares loops (loop width/loop length) $= 0.1$-$1$ with filling factor 1.0. Furthermore, numerical simulations of Yokoyama and Shibata showed that the temperature of evaporated plasma (i.e., $\sim$ "flare temperature" $T$) is a factor of $\sim 3$ lower than the maximum temperature $T_{max}$:

$$T \simeq 1/3 \times T_{max} \simeq 10^7 \Big(\frac{B}{50\text{G}}\Big)^{6/7} \Big(\frac{n_0}{10^9 \text{cm}^{-3}}\Big)^{-1/7} \Big(\frac{L}{10^9 \text{cm}}\Big)^{2/7} \quad \text{K}. \tag{3}$$

(The maximum temperature $T_{max}$ of the reconnection heated loop would correspond to superhot components of flares (Lin et al. 1981, Masuda 1984, Nitta and Yaji 1997).) Since this evaporated plasma has high gas pressure, we have to assume that the magnetic pressure of the reconnected loop must be larger than the gas pressure of evaporated plasma to maintain stable flare loops. From this, it may be reasonable to assume

$$2nkT \simeq \frac{B^2}{8\pi}, \tag{4}$$

as a first approximation. ($k$ is the Boltzmann's constant.) In fact, some previous observations of post flare loops showed that the gas pressure of the flare loop is as high as the inferred magnetic pressure (e.g., Tsuneta 1996).

Eliminating $n$ and $L$ from equations (2)-(4), we find

$$EM \simeq 10^{48} \Big(\frac{B}{50\text{G}}\Big)^{-5} \Big(\frac{n_0}{10^9 \text{cm}^{-3}}\Big)^{3/2} \Big(\frac{T}{10^7 \text{K}}\Big)^{17/2} \quad \text{cm}^{-3}. \tag{5}$$

This is the *EM-T* relation derived by Shibata and Yokoyama (1999), and is hereafter called the *pressure balance scaling law*. We can derive the following equation from above equations

$$EM \simeq 10^{48} \Big(\frac{L}{10^9 \text{cm}}\Big)^{5/3} \Big(\frac{n_0}{10^9 \text{cm}^{-3}}\Big)^{2/3} \Big(\frac{T}{10^7 \text{K}}\Big)^{8/3} \quad \text{cm}^{-3}. \tag{6}$$

The lines of $B =$ constant and $L =$ constant in Figure 2 are drawn based on these two equations (5) and (6). We can also derive the following two equations

$$B = 50 \Big(\frac{EM}{10^{48} \text{cm}^{-3}}\Big)^{-1/5} \Big(\frac{n_0}{10^9 \text{cm}^{-3}}\Big)^{3/10} \Big(\frac{T}{10^7 \text{K}}\Big)^{17/10} \quad \text{G}. \tag{7a}$$

$$L = 10^9 \Big(\frac{EM}{10^{48} \text{cm}^{-3}}\Big)^{3/5} \Big(\frac{n_0}{10^9 \text{cm}^{-3}}\Big)^{-2/5} \Big(\frac{T}{10^7 \text{K}}\Big)^{-8/5} \quad \text{cm}. \tag{7b}$$

These equations will be useful to estimate the magnetic field stregth and length of stellar flare loops which are both difficult to measure with the present observations. Note again that this method depends on only three parameters, the peak temperature ($T$), the peak emission measure ($EM$), and the preflare electron density ($n_0$), and is completely independent of the later evolution of flares after the peak of emission measure. The reason why this method can estimate $B$ and $L$ with fewer observational information than previous methods (van den Oord and Mewe1989, Reale and Micela 1998) is that we assume magnetic reconnection and confinement of evaporated plasma in a reconnected loop (i.e., the balance between evaporated plasma pressure and magnetic pressure of a reconnected loop).

### 2.2. The Enthalpy - Conduction Balance Scaling Law

In deriving equation (5), Shibata and Yokoyama (1999) assumed that the density of flare loops is determined by the balance between gas pressure and magnetic pressure: $2nkT \simeq B^2/8\pi$. However, as they noted in their Discussion section, the density of flare evaporation flow in the initial phase is determined by the balance between enthalpy flux of evaporation flow and conduction flux

$$5nkTC_s \simeq \kappa_0 T^{7/2}/L, \tag{8}$$

where $C_s = (10kT/(3m))^{1/2}$ is the sound speed of the evaporation flow. Then we have

$$n \simeq 10^5 T^2 / L \tag{9a}$$

$$\simeq 10^{10} \Big(\frac{T}{10^7 \text{K}}\Big)^2 \Big(\frac{L}{10^9 \text{cm}}\Big)^{-1} \text{ cm}^{-3}. \tag{9b}$$

This density is a minimum density of evaporating flare loop. Actually, flare loop density gradually increases with time because evaporated plasma acculmulates in the flare loop. If we assume flare loop density is determined by equation (9a), we then find the emission measure $EM$ becomes

$$EM \simeq 10^{47} \Big(\frac{B}{50 \text{G}}\Big)^{-3} \Big(\frac{n_0}{10^9 \text{cm}^{-3}}\Big)^{1/2} \Big(\frac{T}{10^7 \text{K}}\Big)^{15/2} \text{ cm}^{-3}. \tag{10}$$

Hereafter, we call this equation the *enthalpy-conduction balance scaling law*. Eliminating $B$ from equations (3) and (10), we have

$$EM \simeq 10^{47} \Big(\frac{T}{10^7 \text{K}}\Big)^4 \Big(\frac{L}{10^9 \text{cm}}\Big) \text{ cm}^{-3}. \tag{11}$$

In Figure 3, we plot enthalpy-conduction balance scaling law (eq. (10)) for $B = 15, 50$, and 150 G (dashed line) and compare them with pressure balance scaling law (eq. 2)(solid lines). We find that the enthalpy-conduction balance scaling law predicts weaker magnetic field strength than the pressure balance scaling law and actual solar observations. Hence from the viewpoint of magnetic field strength, the pressure balance scaling law fits better with observations. This is also reasonable if we consider the increase in density (beyond the value given by eq. (9a) and (9b)) in the early phase of evolution of a flare loop as discussed in section 4.

### 2.3. Forbidden Region

In the case of the enthalpy-conduction balance (minimum density flare loop), there is no guarantee that gas pressure is smaller than magnetic pressure. From equations (3),(9),(10) the plasma beta of the evaporation flow is estimated to be

$$\beta_{ev} = \frac{2nkT}{B^2/(8\pi)} \simeq 0.3 \Big(\frac{EM}{10^{47}\text{cm}^{-3}}\Big)^{-1/3} \Big(\frac{n_0}{10^9\text{cm}^{-3}}\Big)^{-1/3} \Big(\frac{T}{10^7\text{K}}\Big)^2. \tag{12}$$

If the plasma beta of evaporated flare loop plasma $\beta_{ev}$ becomes larger than unity, the plasma cannot be confined by magnetic pressure. Hence such case does not exist as stable flare loop, and is forbidden in the *EM-T* diagram. This *forbidden region* is defined as the region satisfying the following inequality,

$$EM < 10^{45.5} \Big(\frac{n_0}{10^9\text{cm}^{-3}}\Big)^{-1} \Big(\frac{T}{10^7\text{K}}\Big)^6 \text{cm}^{-3}, \tag{13}$$

which is shown by hatched area in Figure 4. This explains the reason why there is no flare in the lower-right region of the *EM-T* diagram.

### 2.4. General *EM-T* Relation for Arbitrary Heating

The temperature of flares (and also corona; see section 3) is determined by the balance between heating flux ($F$ erg cm$^{-2}$ s$^{-1}$) and conduction cooling flux ($\sim \kappa_0 T^{7/2}/L$),

$$F \simeq \kappa_0 T^{7/2}/L, \tag{14}$$

where $\kappa_0$ ($\simeq 10^{-6}$ cgs) is the Spitzer's thermal conductivity. From this, we obtain

$$T \simeq (FL/\kappa_0)^{2/7} \simeq 10^6 \Big(\frac{F}{10^5 \text{erg cm}^{-2}\text{s}^{-1}}\Big)^{2/7} \Big(\frac{L}{10^{10}\text{cm}}\Big)^{2/7} \text{ K} \tag{15}$$

If the heating is due to magnetic reconnection, the heating flux is

$$F \simeq (B^2/4\pi)V_A \propto B^3. \tag{16}$$

Then, we recover the temperature scaling law (eq. 1)

$$T \propto B^{6/7} L^{2/7}.$$

Using the above equation (15), the emission measure is written as

$$EM \propto B^4 F^{-3} T^{17/2} \quad \text{for pressure balance scaling law} \tag{18a}$$

$$EM \propto F^{-1} T^{15/2} \quad \text{for enthalpy} - \text{conduction scaling law}. \tag{18b}$$

It is now clear from above argument that the observed flare $EM$-$T$ relation corresponds to the relation that heating flux is constant: this is exactly true in the case of enthapy-conduction balance. In the case of pressure-balance, it corresponds to that $B^4 F^{-3}$ or both heating flux and magnetic field strength are constant. In the reconnection heating model, however, heating flux depends only on $B$ so that the $B$ = constant line is the same as the heating flux = constant line.

### 3. Coronal Branch

#### 3.1. Coronal EM-T Scaling Law

If the heating time becomes sufficiently long, the coronal loop becomes dense enough, so that the radiative cooling balances with conduction flux. At that time, we have a steady coronal loop, in which (unknown) heating $F$, conduction flux $\kappa_0 T^{7/2}/L$, and radiative cooling $n^2 \Lambda(T) L$ are all comparable;

$$F \simeq \kappa_0 T^{7/2}/L \simeq n^2 \Lambda(T) L. \tag{19}$$

Here the radiative loss function $\Lambda(T)$ is approximated to

$$\Lambda(T) \simeq 3 \times 10^{-23} (T/10^7 \text{K})^{-1/2} \quad \text{for } T < 10^7 \text{ K}, \tag{20a}$$

$$\Lambda(T) \simeq 3 \times 10^{-23} (T/10^7 \text{K})^{1/2} \quad \text{for } T > 10^7 \text{ K}. \tag{20b}$$

The equation (19), (20a) and (20b) are reduced to

$$n \simeq 10^{6.5} T^2 / L \quad \text{for } T < 10^7 \text{ K} \tag{21a}$$

$$n \simeq 10^{10} T^{1.5} / L \quad \text{for } T > 10^7 \text{ K}. \tag{21b}$$

Note that the equation (21a) is basically the same as the Rosner-Tucker-Vaiana (1978)'s scaling law ($T \propto (pL)^{1/3}$). It is interesting to note that the equation (21a) has the same dependence on $T$ and $L$ with the equation (9a) for the enthalpy flux - conduction balance.

Then, the *EM-T* relation becomes

$$EM \simeq 10^{47} \left(\frac{T}{10^6 \text{K}}\right)^{15/2} \left(\frac{F}{10^5 \text{erg/cm}^2/\text{s}}\right)^{-1} \text{ cm}^{-3} \text{ for } T < 10^7 \text{ K}, \quad (22a)$$

$$EM \simeq 10^{52.5} \left(\frac{T}{10^7 \text{K}}\right)^{13/2} \left(\frac{F}{10^7 \text{erg/cm}^2/\text{s}}\right)^{-1} \text{ cm}^{-3} \text{ for } T > 10^7 \text{ K}. \quad (22b)$$

Eliminating $F$ from (22a) using (19), we obtain

$$EM \simeq 10^{47} \left(\frac{T}{10^6 \text{K}}\right)^4 \left(\frac{L}{10^{10} \text{cm}}\right) \text{ cm}^{-3} \text{ for } T < 10^7 \text{ K}, \quad (23a)$$

$$EM \simeq 10^{51} \left(\frac{T}{10^7 \text{K}}\right)^3 \left(\frac{L}{10^{10} \text{cm}}\right) \text{ cm}^{-3} \text{ for } T > 10^7 \text{ K}. \quad (23b)$$

### 3.2. Comparison with Observations and Effect of Filling Factor

Figure 5 shows this coronal branch (eq.(23a) and (23b)) in the *EM-T* diagram as well as observed *EM-T* relations of solar active regions (Yashiro 2000). We see that the theoretical *EM-T* relation for solar corona fits well with observations. However, the heating flux necessary for the heating of active region corona becomes $10^7 - 10^8$ erg cm$^{-2}$ s$^{-1}$, which is larger than the heating flux required for the heating of average active region corona ($\sim 10^6 - 10^7$ erg cm$^{-2}$ s$^{-1}$, e.g., Withbroe and Noyes 1977). This is probably because of the effect of the filling factor. In deriving above relations, we implicitly assumed that the filling factor of the corona with $n$ and $T$ in the volume of $L^3$ is unity. If we take into account the effect of the filling factor $f$, the emission measure is written as

$$EM = fn^2 L^3, \quad (24)$$

so that the coronal branch scaling law becomes

$$EM \simeq 10^{46} \left(\frac{f}{0.1}\right) \left(\frac{T}{10^6 \text{K}}\right)^{15/2} \left(\frac{F}{10^5 \text{erg/cm}^2/\text{s}}\right)^{-1} \text{ cm}^{-3} \text{ for } T < 10^7 \text{ K}, \quad (25a)$$

$$EM \simeq 10^{51.5} \left(\frac{f}{0.1}\right) \left(\frac{T}{10^7 \text{K}}\right)^{13/2} \left(\frac{F}{10^7 \text{erg/cm}^2/\text{s}}\right)^{-1} \text{ cm}^{-3} \text{ for } T > 10^7 \text{ K}. \quad (25b)$$

If we assume $f = 0.1$, the heating flux necessary for the heating of average active region corona becomes consistent with observations (see Fig. 5). It is also interesting to note that the theory predict that the coronal branch for protostars is closer to the flare branch than in the solar case.

Figures 5 show curves of constant flare loop length and comparison with stellar observations (Hamaguchi et al. 2000, Hamaguchi 2001, Ozawa et al. 1999, Ozawa 2000,

Imanishi et al. 2001) for $f = 0.1$. Stellar coronal data (non-flare data) are distributed in wider area compared with solar active region data. This might mean that the coronal heating flux in stellar corona has wider distribution than in the solar corona. Alternatively, this may simply be a result of "stellar microflares". That is, even in non-flare stellar corona, there would be a lot of "microflares" as in the solar corona. Such "stellar microflares" may be seen in these data. In fact, in the case of the solar coronal data analysis, microflare data are carefully removed from the "solar corona" data (Yashiro 2000), and the combination of the solar corona and solar microflares show wide distribution similar to that of stellar non-flare data. If this interpretation is right, the data in the left-most side of stellar non-flare data (with $EM \sim 10^{51} - 10^{53} \mathrm{cm}^{-3}, T \sim 1 - 2 \times 10^7 \mathrm{K}$) correspond to true (non-microflare) stellar corona, and suggest that the heating flux $\sim 10^7 - 10^8$ erg cm$^{-2}$ s$^{-1}$ which is an order of magnitude larger than the solar coronal heating flux, and the stellar coronal loop length $\sim 10^{11} - 10^{12}$ cm is also one-two orders of magnitude larger than the size of solar active region corona if the filling factor $= 0.1$.

## 4. Discussion

### 4.1. Flare Evolution Track

In section 1, we have noted that the $EM$-$T$ diagram is similar to HR diagram for stars. In the case of HR diagram, we can plot stellar evolution track on it. How about flare evolution in the $EM$-$T$ diagram ? Yes, we can plot flare evolution track in the $EM$-$T$ diagram as discussed below.

The flare evolution is often discussed in the $n$-$T$ diagram (Fig. 6), which is obtained from one-dimensional hydrodynamic numerical simulations of flare loops (e.g., Jakimiec et al. 1992, Sylwester 1996, for a review). In this case, when a sudden flare energy release occurs, the temperature of flare plasma increases until the conduction front reaches the top of the chromosphere. After that, the temperature becomes constant, $T_{max}$ (eq. 1), and the evaporation starts with density $n_{ev}$ (eq. 9) determined by the enthalpy-conduction balance. The flare loop density gradually increases owing to evaporation as long as the flare heating continues. If the heating time is long enough, the loop density becomes comparable to the density ($n_{RTV}$ (eq. 21)) determined by the radiation-conduction balance. (This is equivalent to the Rosner-Tucker-Vaiana (1978)'s scaling law for a coronal loop.) Usually, the flare heating time is short for a particular loop (Hori et al. 1997) so that the flare density does not reach $n_{RTV}$, and the temperature as well as density of flare loop gradually decrease with time, reaching initial state (corona) eventually. This is a full story of *flare evolution track* in the $n$-$T$ diagram.

We can plot similar *flare evolution track* on the $EM$-$T$ diagram and for a flare loop with fixed loop length $L$ (Fig. 7). From above discussion, we see that one reason of scatter of various flares ("main sequence" flares) on $EM$-$T$ diagram may be due to difference of flare heating time which leads to different flare density (thus $EM$).

### 4.2. Nanoflare

If the flare loop length is smaller than $\sim 10^7 - 10^8$ cm, the top of the flare loop is lower than the height of the transition region. In this case, the loop density is so large that it is difficult to produce high temperature plasmas, i.e., flares. Hence for loops shorter than $10^7$ cm, there is no "flares" (i.e., we have another forbidden region). Along $L = 10^7$ cm, $EM \simeq 10^{41} - 10^{44}$ cm$^{-3}$ for $T \simeq 10^5 - 10^6$ K. These flares correspond to nanoflares observed with the EUV imaging telescope such as SOHO/EIT and TRACE.

Using EUV data of nanoflares observed with SOHO/EIT and TRACE, and soft X-ray data of microflares and flares observed with Yohkoh/SXT, Aschwanden (2000) empirically obtained $EM \propto T^7$. We plotted his data on Fig. 4. We see that nanoflare is on the $L = $ constant $= 10^7$ cm line ($EM \propto T^{8/3}$), and microflares - flares are on the $EM \propto T^{15/2}$ or $T^{17/2}$ line. Hence we suggest that Aschwanden (2000)'s $EM \propto T^7$ is explained by a combination of $EM \propto T^{8/3}$ line for nanoflares ($T < 10^6$ K) and $EM \propto T^{15/2}$ or $T^{17/2}$ line for microflares ($T > 10^6$ K).

### 4.3. Total Energy Released by Flares and Comparison with Recent Stellar Flare Observations

As emphasized in previous sections, the $EM$-$T$ diagram is useful to estimate unobservable physical quantities such as magnetic field strength, flare loop length etc. from observed three parameters $EM$, $T$, and pre-flare density $n_0$ (see Appendix 1). Among these three quantities, it may be difficult to estimate the pre-flare density. In that case, we shall assume $n_0 = 10^9$ cm$^{-3}$ which is a typical value of the solar active region coronal density (Yashiro 2000). Then we can estimate the total energy released by a flare, which is simply assumed to be the total magnetic energy included in a flare loop

$$E_{mag} = \frac{B^2}{8\pi} L^3, \tag{26}$$

which is comparable to the total thermal energy content of a flare loop

$$E_{th} = 3nkTL^3 = \frac{3}{2} E_{mag} \tag{27}$$

by our pressure balance assumption. Using equations (26) and those in Appendix 1, we find

$$EM = 10^{48} \Big(\frac{E_{mag}}{10^{29}\text{erg}}\Big)^{5/7} \Big(\frac{T}{10^7\text{K}}\Big) \Big(\frac{n_0}{10^9\text{cm}^{-3}}\Big)^{3/7} \text{ cm}^{-3}. \tag{28}$$

This relation is plotted in Figure 8 for $10^{26}, 10^{29}, 10^{32}, 10^{35}, 10^{38}$ erg. We immediately find that energy ranges of solar microflares and solar flares are $10^{26} - 10^{29}$ erg, $10^{29} - 10^{32}$ erg, respectively, both of which are consistent with previous estimate. We also plotted new observational data of stellar flares from ASCA (Hamaguchi 2001, Ozawa 2000) and Chandra (Imanishi et al. 2001). It is found that stellar flare energy ranges from $10^{34}$ erg to $10^{38}$ erg, if the pre-flare coronal density is $\sim 10^9$ cm$^{-3}$. These values are again comparable to previous estimate of stellar flare energy using other methods.

Finally, we should emphasize that our method is based only on observed peak temperature, peak emission measure, and pre-flare coronal density. We do not have to measure detailed time evolution of flares. This is in contrast with other method such as the method using the flare decay time (e.g., van den Oord and Mewe 1989, Reale and Micela 1998).

### 4.4. Comment on the Stellar Flare Loop Length

Recently, using a hydrodynamic model (Reale and Micela 1998), Favata et al. (2001) re-analyzed large flaring events on pre-main sequence stars, and found that the size of the flaring regions is much smaller ($L < R_*$) than previous estimate ($L > R_*$) based on the quasi-static model (van den Oord and Mewe 1989). Here $R_*$ is stellar radius of order of $10^{11}$ cm. Our results in Figure 2 predict that many stellar flares have loop length larger than $10^{11}$ cm. Are these results completely inconsistent with Favata et al.'s results ? Two comments should be made for this kind of comparison. First, Figure 2 is based on the assumption that the preflare coronal density is $10^9$ cm$^{-3}$. If $n_0 = 10^{10}$ cm$^{-3}$ which can often occur in very active solar active regions, the predicted loop length based on equation (7) would decrease by a factor of $10^{2/5} \simeq 2.5$. Second, if the flare temperature is a factor 2 - 2.5 larger than the original estimate as discussed in Favata et al. (2001), the predicted loop length would be smaller by a factor of $2^{-8/5} - 2.5^{-8/5} \simeq 3 - 4$ (see equation (7)). In fact, in the case of faint stellar flares, the long exposure time may reduce the observed peak temperature than the real one. In either case, the loop length can decrease by a factor of $\sim 3$, which may fit with the results by Favata et al. (2001). In order to clarify whether these effects (preflare density and true flare temperature) are really important or not, more detailed study will be necessary, which take into account the pass band effect of observing

instruments. [1]

## 5. Conclusions

In this paper, we proposed that the $EM$-$T$ diagram for solar/stellar flares is similar to HR diagram for stars. The $EM$-$T$ diagram shows the following properties:

(1) It shows universal correlation $EM \propto T^{17/2}$ (for pressure balance loops) or $T^{15/2}$ (for enthalpy-conduction balance loops) for the flare peak $EM$ and $T$. The former agrees with observations better than the latter. These may be called *main sequence flares*. Physically, the main sequence flare corresponds to the flare with magnetic field strength = constant or equivalently heating flux = constant.

(2) It has the *forbidden region* which correspond to the region where the gas pressure of evaporated plasma exceeds magnetic pressure of a flare loop so that a stable flare loop cannot exist.

(3) It shows the *coronal branch* with $EM \propto T^{15/2}$ for $T < 10^7$ K and $EM \propto T^{13/2}$ for $T > 10^7$ K. This corresponds to the Rosner-Tucker-Vaiana (1978)'s scaling law.

(4) There is another *forbidden region* determined by the length of a flare loop; a lower limit of flare loop is $\sim 10^7$ cm. Those small flares on $L$ = constant line have low temperature ($T < 10^6$ K), thus explaining EUV nanoflares. This may be called the *nanoflare branch*.

(5) We can plot the *flare evolution track* on the $EM$-$T$ diagram. A flare evolves from the coronal branch to main sequence flares, and then returns to the coronal branch eventually.

(6) Using the $EM$-$T$ diagram, we can estimate unobservable physical quantities, such as the magnetic field strength, flare loop length, total flare energy, etc., using only three observed quantities, the peak temperature $T$, peak emission measure $EM$, and pre-flare coronal density $n_0$. We do not need observations of detailed flare evolution. Hence even with limited data around the flare peak intensity and pre-flare phase, we can estimate the magnetic field strength, flare loop length, total flare energy, and etc.

The authors would like to thank S. Yashiro for allowing us to use his solar coronal data on $EM$-$T$ relation before publication. They also thank K. Hamaguchi, K. Imanishi,

---

[1] For more detailed studies, the effects of filling factor and unseen (out of band) emission measure should also be considered, though both effects would increase the predicted flare loop lengths (see eq. (A14)).

K. Koyama, F. Nagase, H. Ozawa, Y. Tsuboi, T. Watanabe for useful discussion on stellar flares and corona. This work is supported in part by the JSPS grant for Japan-US Collaboration in Scientific Research.

## Appendix 1: Useful formula

Using the *EM-T* diagram, we can estimate various physical quantities of solar and stellar flares. Here we summarize useful formula for estimating various physical quantities on the basis of three observables $(EM, T, n_0)$. Here, it should be noted that $n_0$ is the electron density in pre-flare corona (i.e., pre-evaporation corona), and is not equal to the electron density of a flare loop. It is also noted that $n_0$ is not necessarily well observed. In the case of the pre-flare solar corona in active regions, Yohkoh observations revealed $n_0 \sim 10^9$ cm$^{-3}$ on average (e.g., Yashiro 2000, Yashiro and Shibata 2001). Hence we often assume $n_0 = 10^9$ cm$^{-3}$ when discussing actual application in this paper, though we leave the dependence on $n_0$ in many formula.

The basic quantities, magnetic field strength $B$, flare loop length $L$, flare loop electron density $n$ (after evaporation), can be all derived from $(EM, T, n_0)$:

$$B_{50} = EM_{48}^{-1/5} T_7^{17/10} n_{09}^{3/10}, \quad (A1)$$

$$L_9 = EM_{48}^{3/5} T_7^{-8/5} n_{09}^{-2/5}, \quad (A2)$$

$$n_9 = 10^{1.5} EM_{48}^{-2/5} T_7^{12/5} n_{09}^{3/5}. \quad (A3)$$

Here, $B_{50} = B/(50 \text{ G})$, $EM_{48} = EM/(10^{48} \text{ cm}^{-3})$, $T_7 = T/(10^7 \text{ K})$, $n_{09} = n_0/(10^9 \text{ cm}^{-3})$, $L_9 = L/(10^9 \text{ cm})$, and $n_9 = n/(10^9 \text{ cm}^{-3})$.

Using above quantities, for *pre-evaporation corona*, we can calculate the Alfven time (= reconnection heating time) $\tau_A$, conduction cooling time $\tau_c$, radiative cooling time $\tau_r$:

$$\tau_A = L/V_A = L(4\pi nm)^{1/2}/B \simeq 3 \; T_7^{-33/10} EM_{48}^{4/5} n_{09}^{-1/5} \text{ sec}, \quad (A4)$$

$$\tau_c = \frac{3nkT}{\kappa_0 T^{7/2}/L^2} \simeq 1.4 \; T_7^{-57/10} EM_{48}^{6/5} n_{09}^{1/5} \text{ sec}, \quad (A5)$$

$$\tau_r = \frac{3nkT}{n^2 \Lambda(T)} \simeq 1.4 \times 10^5 \; T_7^{3/2} n_{09}^{-1} \text{ sec} \quad \text{for } T < 10^7 \text{K}, \quad (A6a)$$

$$\tau_r \simeq 1.4 \times 10^5 \; T_7^{1/2} n_{09}^{-1} \text{ sec} \quad \text{for } T > 10^7 \text{K}. \quad (A6b)$$

Here, $\Lambda(T)$ is given by equations (20a) and (20b).

## Appendix 2: Limit of Application for Flare Temperature Formula

In deriving the flare temperature formula (eq. (1)), we implicitly assumed that the conduction cooling time $\tau_c$ (eq. (A4)) is shorter than the reconnection heating time $\tau_A$ (eq. (A5)). Here we shall derive the condition that this inequality

$$\tau_c < \tau_A \qquad (A7)$$

is satisfied. Using equations (A4) and (A5), we find that inequality becomes

$$EM_{48} < 10^{5/2} \; T_7{}^6 n_{09}{}^{-1}. \qquad (A8)$$

It is seen from Figure 9 that most of data (for $B = 50\text{-}150$ G) for solar and stellar flares satisfies this condition. In the case of stellar flares, some of data lie above the critical line. If the pre-flare coronal density is higher, say $10^{10}$ cm$^{-3}$, more data become marginal.

Above this line (Fig. 9), the flare maximum temperature is determined in situ by slow shock heating or ohmic heating in the diffusion region, if the radiative cooling time $\tau_r$ is longer than the conduction time $\tau_c$. That is, in order-of-magnitude estimate, the flare temperature is determined by the energy (pressure) balance $p \sim B^2/8\pi$ and is written as

$$T_{flare} \sim B^2/(16\pi nk). \qquad (A9)$$

## Appendix 3: Condition for Occurrence of Chromospheric Evaporation

If the radiation cooling time $\tau_r$ becomes comparable to or shorter than the conduction cooling time $\tau_c$ in the reconnection-heated (but pre-evaporation) flare plasmas, the evaporation cannot occur any more. In this case, heating flux is balanced with in-situ radiative cooling. Hence the condition for chromospheric evaporation is

$$\tau_c < \tau_r. \qquad (A10)$$

Using equations (A5), (A6a) and (A6b), we find that this inequality is rewritten as

$$EM_{48} < 10^{4.2} T_7{}^6 n_{09}{}^{-1} \quad \text{for } T < 10^7 \text{ K}, \qquad (A11a)$$

$$EM_{48} < 10^{4.2} T_7{}^{31/6} n_{09}{}^{-1} \quad \text{for } T > 10^7 \text{ K}. \qquad (A11b)$$

The line $\tau_c = \tau_r$ is shown in Figure 10. It is seen from Figure 10 that the main flare sequence (for $B = 50\text{-}150$ G) for solar and stellar flares satisfies this condition.

– 16 –## Appendix 4: Effect of Filling Factor on Flare EM-T Scaling Law

In the previous paper as well as in section 2 of this paper, we implicitly assumed that the flare loop with volume $L^3$ is filled with hot dense plasma with single temperature $T$ and density $n$. However, actually this may not be true and we have to consider the effect of the filling factor $f$. In this case, we should write the emission measure including the effect of the filling factor

$$EM = fn^2L^3,$$

which is the same as eq. (24). Then the flare $EM$-$T$ relation for the pressure balance scaling law becomes

$$EM \simeq 10^{47}\Big(\frac{f}{0.1}\Big)\Big(\frac{B}{50\mathrm{G}}\Big)^{-5}\Big(\frac{n_0}{10^9\mathrm{cm}^{-3}}\Big)^{3/2}\Big(\frac{T}{10^7\mathrm{K}}\Big)^{17/2} \quad \mathrm{cm}^{-3}. \quad (A12)$$

Hence the effect of the filling factor is only to decrease the emission measure by that factor. The physical quantities derived from observed $(EM, T, n_0)$ with the effect of the filling factor are now written as

$$B_{50} = f_{0.1}^{1/5} EM_{47}^{-1/5} T_7^{17/10} n_{09}^{3/10}, \quad (A13)$$

$$L_9 = f_{0.1}^{-3/5} EM_{47}^{3/5} T_7^{-8/5} n_{09}^{-2/5}, \quad (A14)$$

$$n_9 = 10^{1.5} f_{0.1}^{2/5} EM_{47}^{-2/5} T_7^{12/5} n_{09}^{3/5}, \quad (A15)$$

where $f_{0.1} = f/(0.1)$. Unlike the coronal case (section 3.2), there is no strong reason that the filling factor must be $\sim 0.1$ for solar and stellar flares.

- 19 -Figure Caption

Figure 1

The log-log plot of emission measure vs. electron temperature of solar flares (from Feldman et al. 1995), solar microflares observed by Yohkoh SXT (from Shimizu 1995), four stellar flares (asterisks, from Feldman et al. 1995), a protostellar flare (diamond, class 1 protostar far IR source R1 in the R CrA cloud, from Koyama et al. 1996), a T-Tauri stellar flare (diamond, weaklined T-Tauri star V773 Tau, from Tsuboi et al. 1998), and a stellar flare on AB Dor (K0 IV ZAMS single star) by BeppoSax (cross, Pallavicini 2001).

Figure 2

The theoretical $EM$-$T$ relations (pressure balance scaling law: eq. (5)) overlaid on the observed $EM$-$T$ relation based on recent observations with ASCA (Koyama et al. 1996, Tsuboi et al. 1997, Hamaguchi et al. 2000, Hamaguchi 2001, Ozawa et al. 1999, Ozawa 2000) and Chandra (Imanishi et al. 2001). Four theoretical curves are plotted for different magnetic field strengths $B$ = 5, 15, 50, and 150 G. The flare loop length $L$ = constant curves (eq. 6) are also plotted with dash-dotted lines for $L = 10^8, 10^{10}, 10^{12}$ cm. It is interesting to note that all the observed stellar flare data points lie below the flare loop length $L = 10^{12}$ cm line, which may simply represent the upper limit of the flare loop size in stars.

Figure 3

The $EM$-$T$ relation for the enthalpy-conduction balance scaling law (eq. 10; $EM \propto T^{15/2}$) (solid lines) which is compared with that for the pressure balance scaling law (eq. 5; $EM \propto T^{17/2}$ ) (dashed lines) as well as observed data (Fig. 1). The three cases are plotted for different field strength, $B$ = 15, 50, and 150 G. Note that the observed data agree better with the pressure balance scaling law than the enthalpy-conduction balance scaling law.

Figure 4

The forbidden regions, $\beta_{ev} > 1$ ($EM < 10^{45.5}T_7{}^6$ cm$^{-3}$, see eq. 12) and flare loop height $< 10^7$ cm ($EM < 10^{44.7}T_7{}^{8/3}$ cm$^{-3}$, see eq. 6 and section 4.2) are plotted in the $EM$-$T$ diagram. Here, $T_7 = T/10^7$K. The branch around $T < 10^6$ K corresponds to nanoflare; the point A corresponds to EUV nanoflare (Aschwanden 2000) and the point D corresponds to explosive events (Dere et al. 1989).

Figure 5

Coronal branch (eq. 22a and 22b) is shown with dash-dotted lines in the $EM$-$T$ diagram for three cases; heating flux $F = 10^6, 10^7, 10^8$ erg cm$^{-2}$ s$^{-1}$. The filling factor $f = 0.1$ is assumed. The data for solar corona (active region; Yashiro 2000) are indicated with hatched area around $T \sim 2 - 3 \times 10^6$ K. Coronal loop length $L_c$ is denoted with solid lines for $L_c = 10^8, 10^9, 10^{10}, 10^{11}, 10^{12}$ cm on the coronal $EM$-$T$ scaling law. Recent observations of non-flare data with ASCA (cross: Herbig Be (Hamaguchi et al. 2000); triangle : T-Tauri (Ozawa 2000)) and Chandra (square: Class I, II, III (Imanishi et al. 2001)) are also plotted. The thick solid line is the theoretical correlation line for flare sequence ($EM \propto T^{17/2}$ for $B = 50$G, see Fig. 2).

Figure 6

The flare evolution track in the $n$-$T$ diagram (e.g., Jakimiec et al. 1992). When a flare loop start to be heated, the temperature increases rapidly to the maximum value until the conduction front reaches the top of the chromosphere. After that, the increase in temperature stops because of start of evaporation (i.e., conduction balances with enthalpy flux). The flare density increases due to evaporation as long as the flare heating continues. Usually, the flare heating time is short so that the maximum density is lower than the Rosner-Tucker-Vaiana (RTV) (1978)'s steady coronal value in which conduction time ($\tau_c$) is comparable to radiative cooling time ($\tau_r$). After that, the temperature and density gradually decrease and eventually return to the initial pre-flare coronal value.

Figure 7

The flare evolution track in the $EM$-$T$ diagram. Only the early phase (until the peak emission measure) is shown for four different loop length cases. In actual observations (especially in stellar flare/corona observations), the data points on this evolution track (i.e. early phase) as well as those in the decay phase would be observed. The scatter of data points in actual observations may result from such evolutionary effects.

Figure 8

The lines with total flare energy $=$ constant (eq. 27) are shown in the $EM$-$T$ diagram. The recent stellar flare data (Hamaguchi 2001, Hamaguchi et al. 2000, Ozawa et al. 1999, Ozawa 2000, Imanishi et al. 2001) are also indicated.

Figure 9

The critical line ($\tau_A = \tau_c$; eq (A8)) above which the temperature scaling law (eq. 1) is no longer valid. The observed data for solar and stellar flares are also shown.

Figure 10

The critical line ($\tau_c = \tau_r$; eqs. (A11a) and (A11b)) above which the chromospheric evaporation does no longer occur. The $\tau_A = \tau_c$ is also shown with observed data for solar and stellar flares.

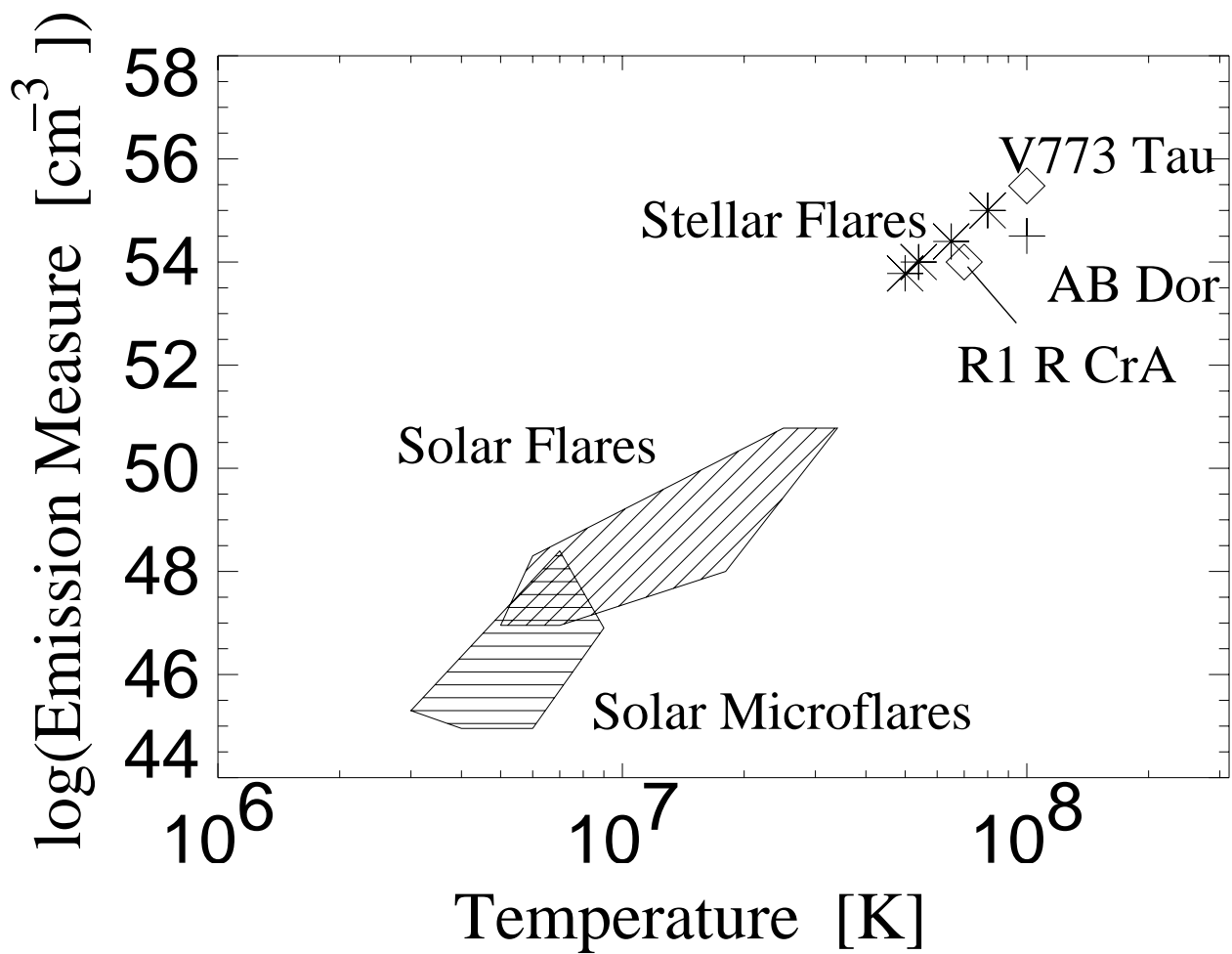

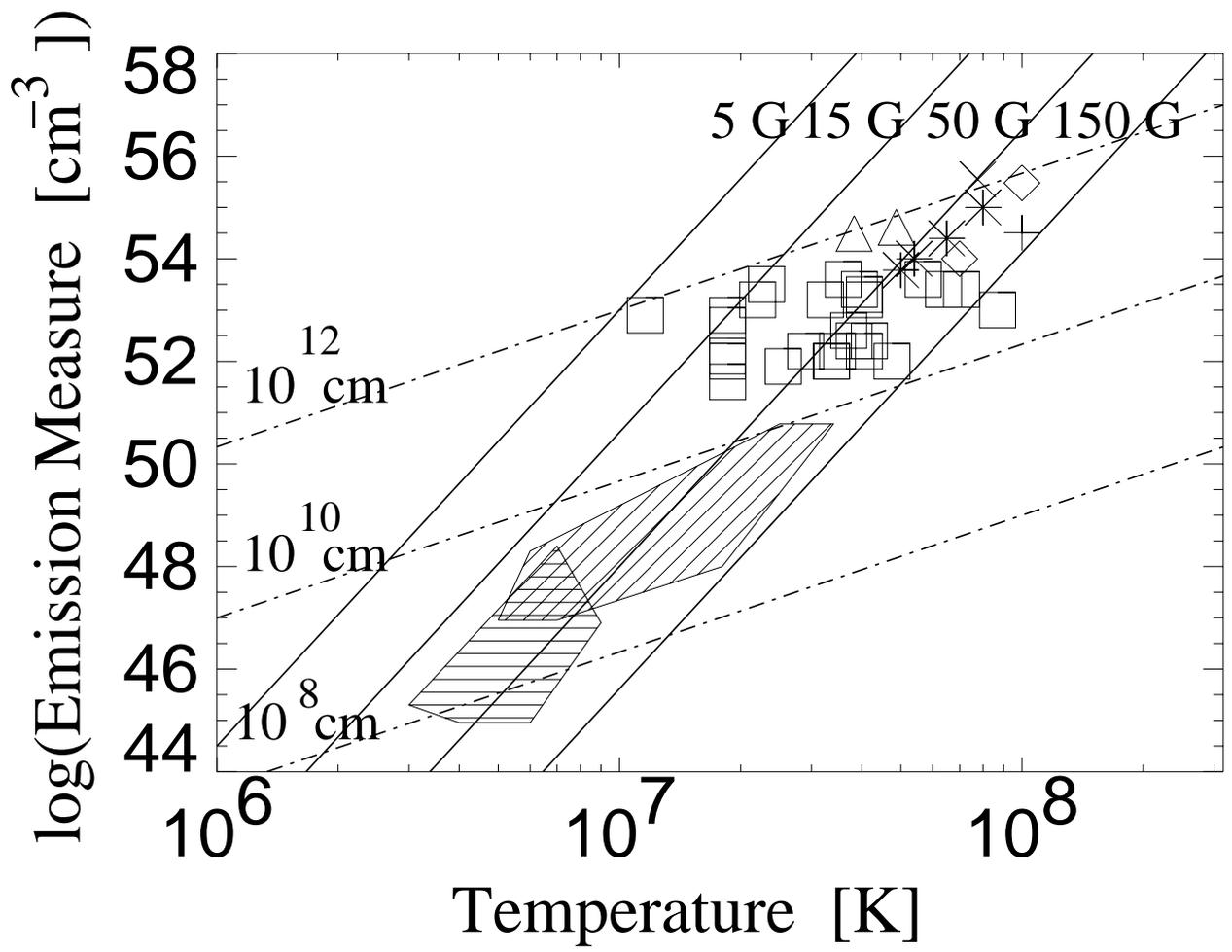

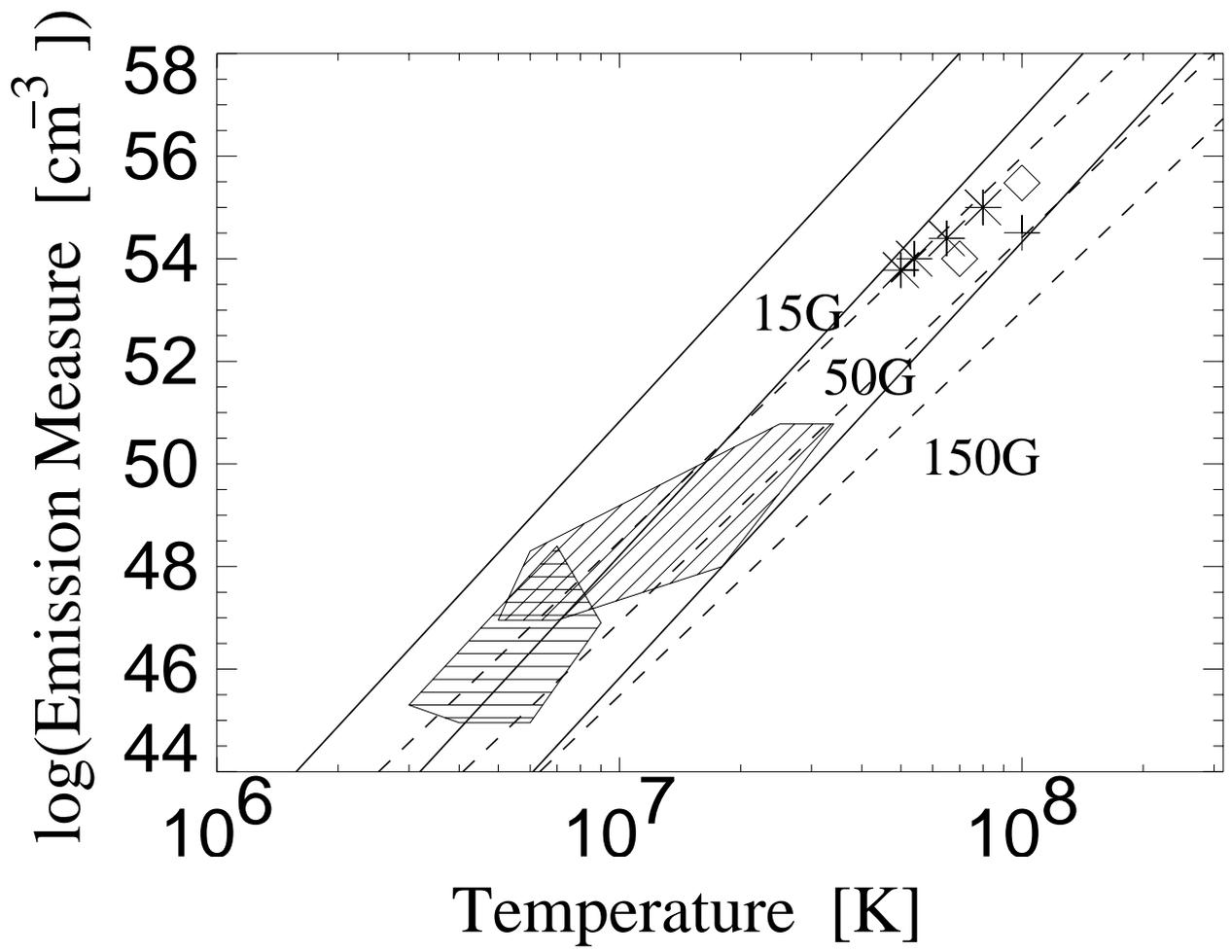

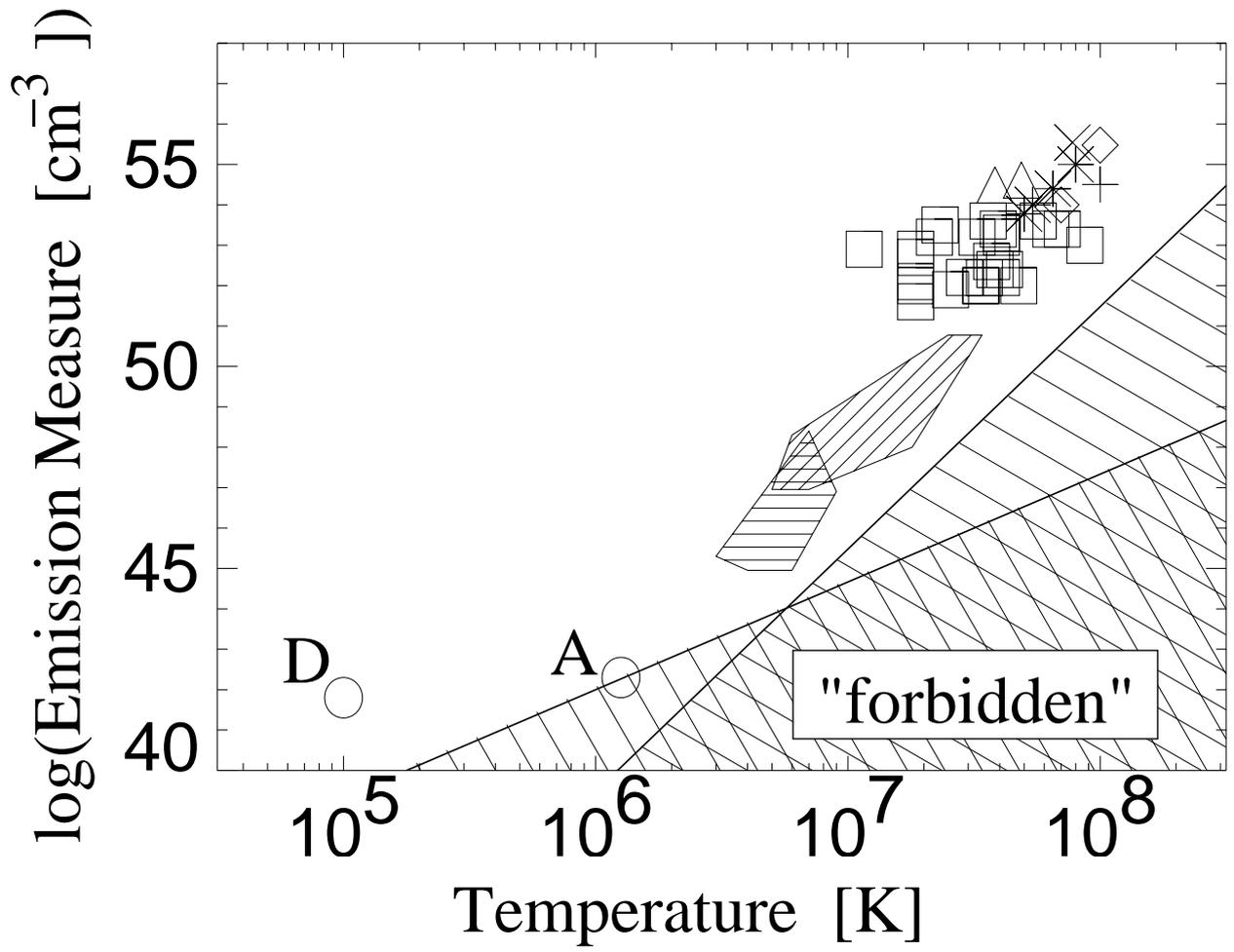

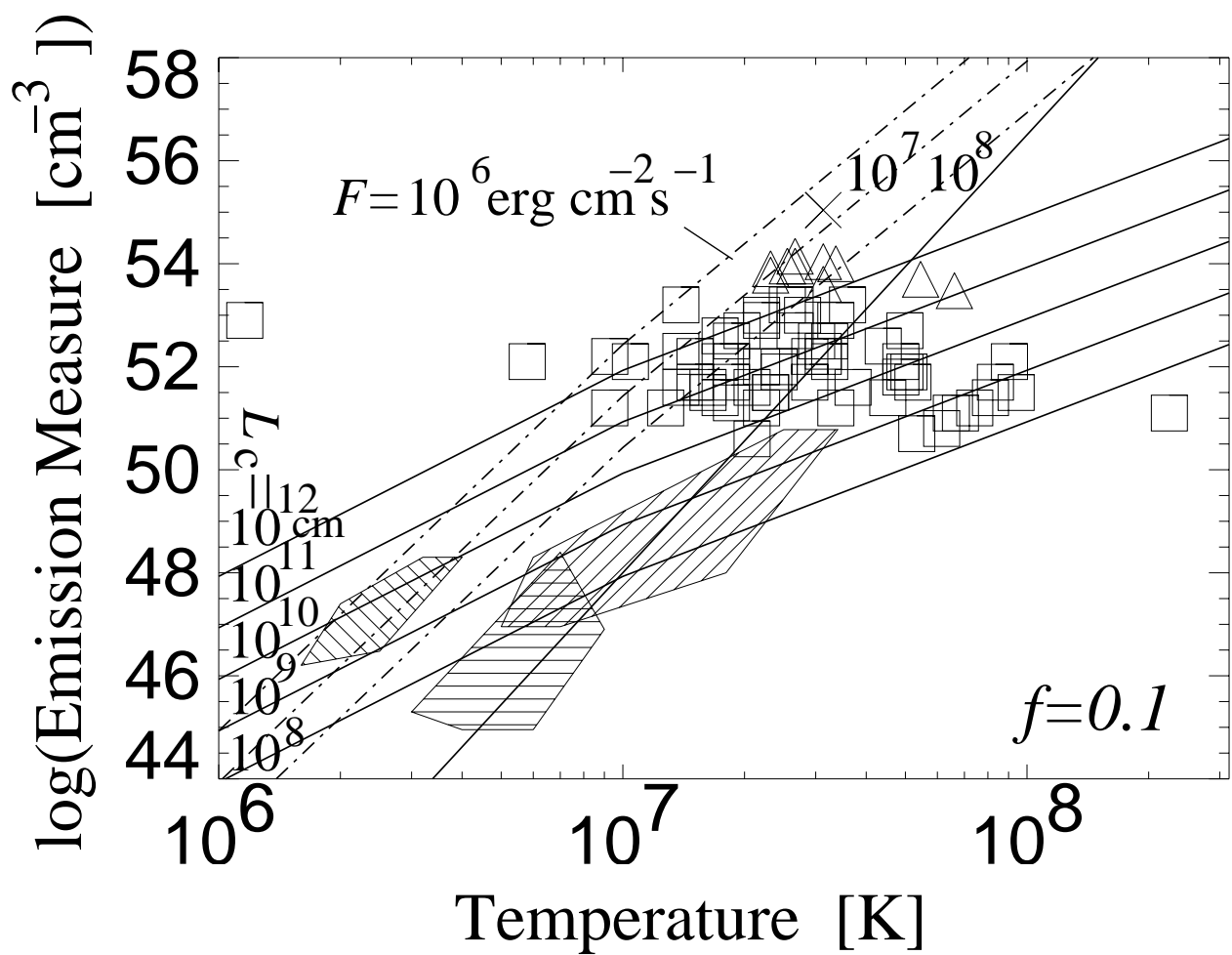

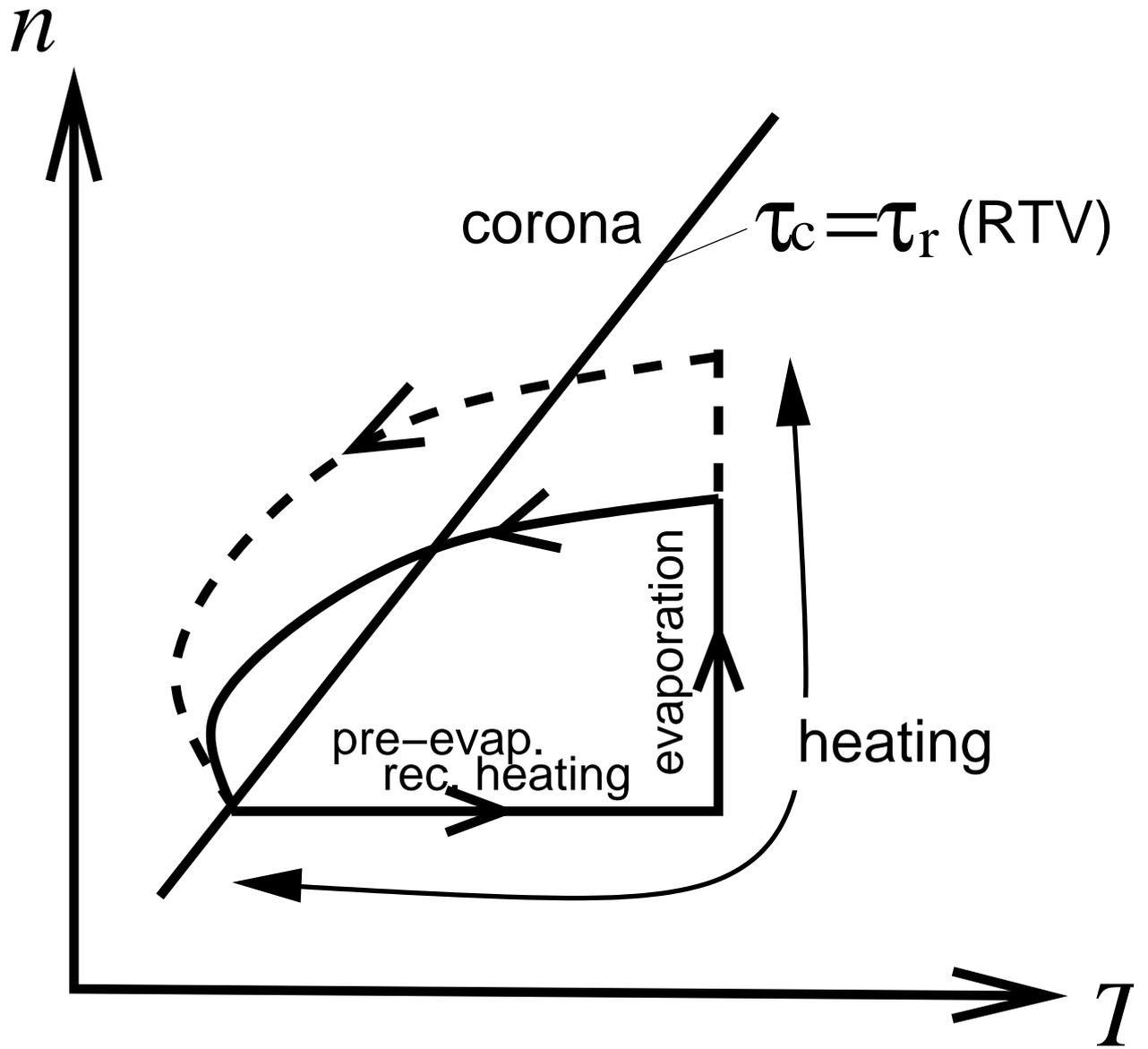

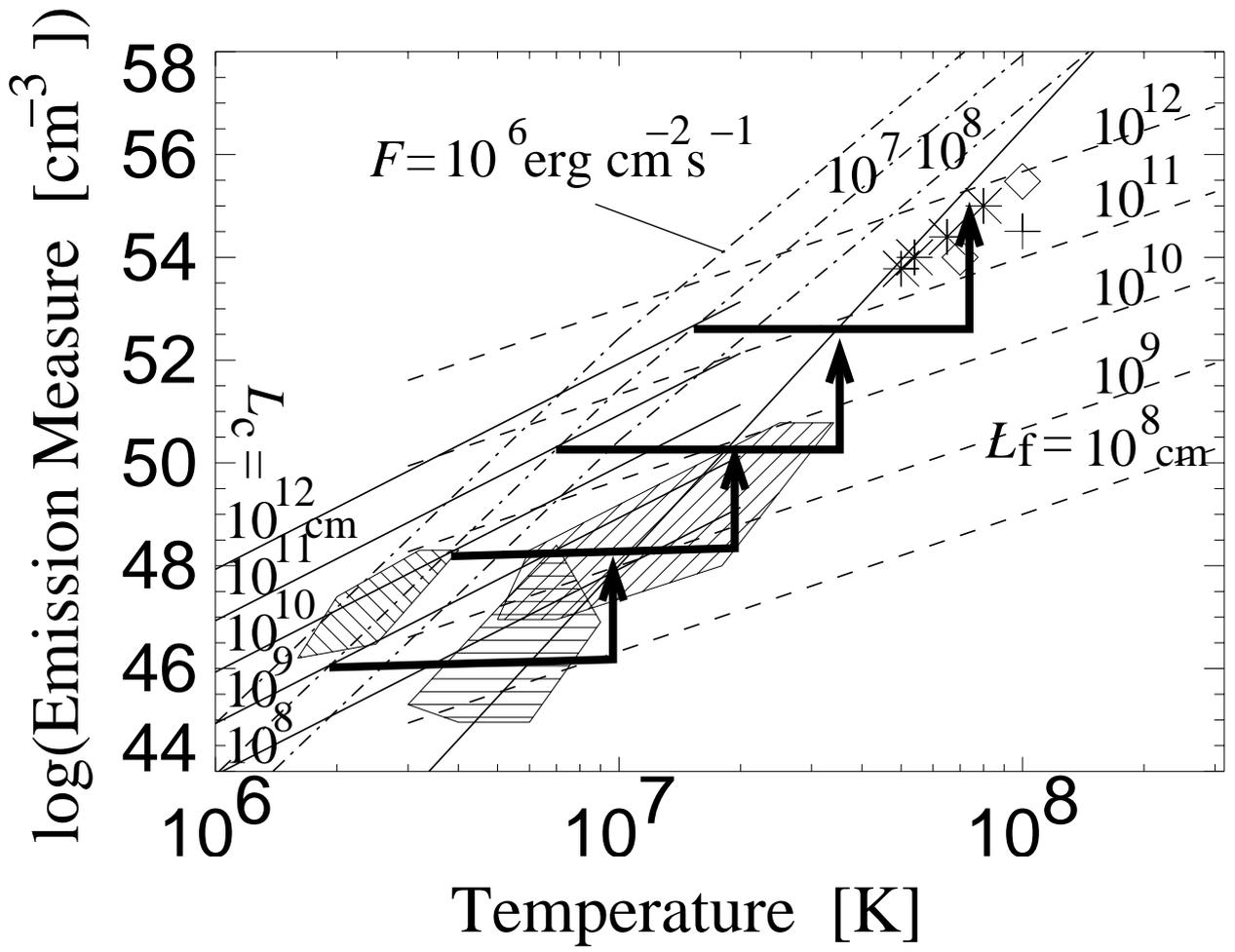

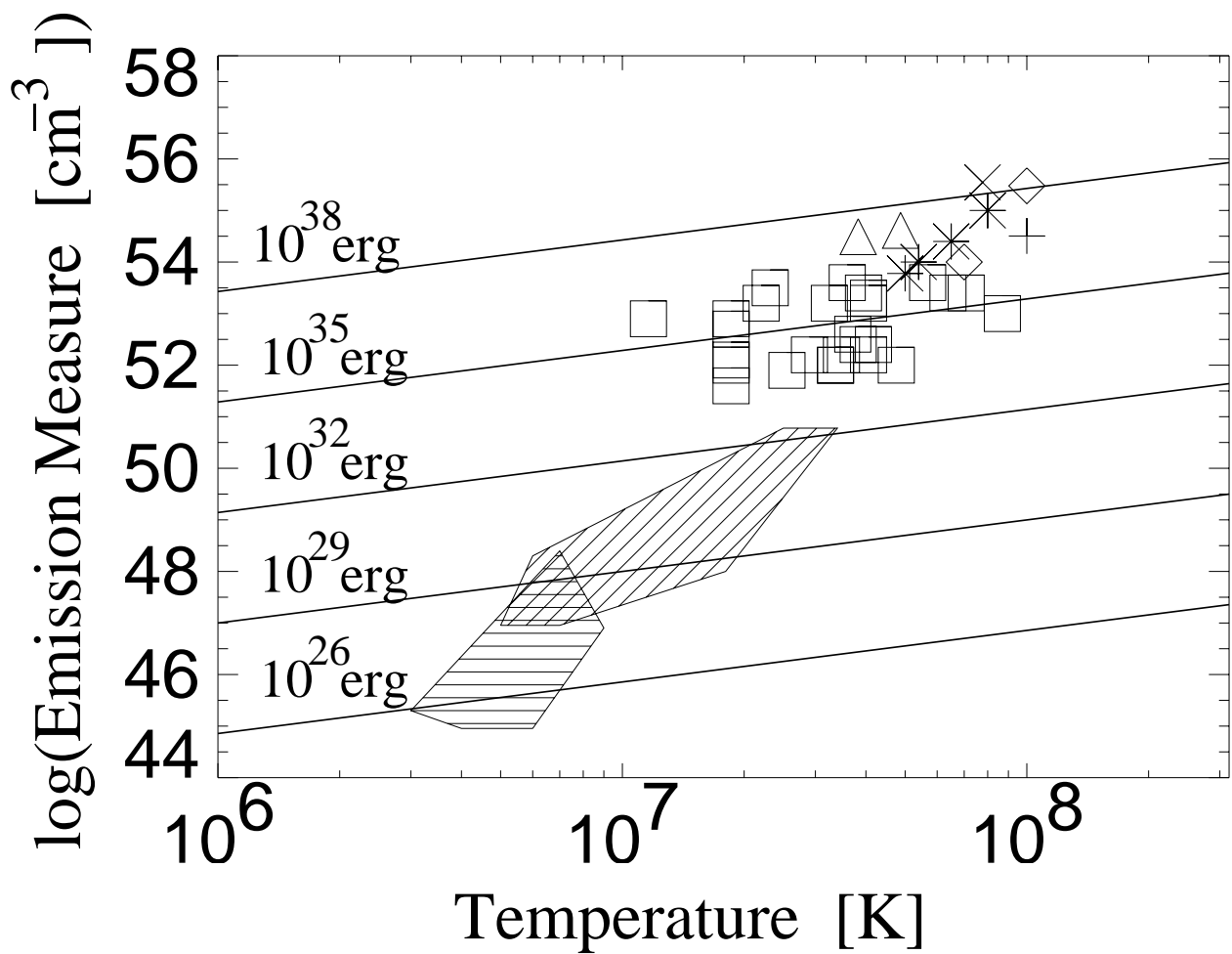

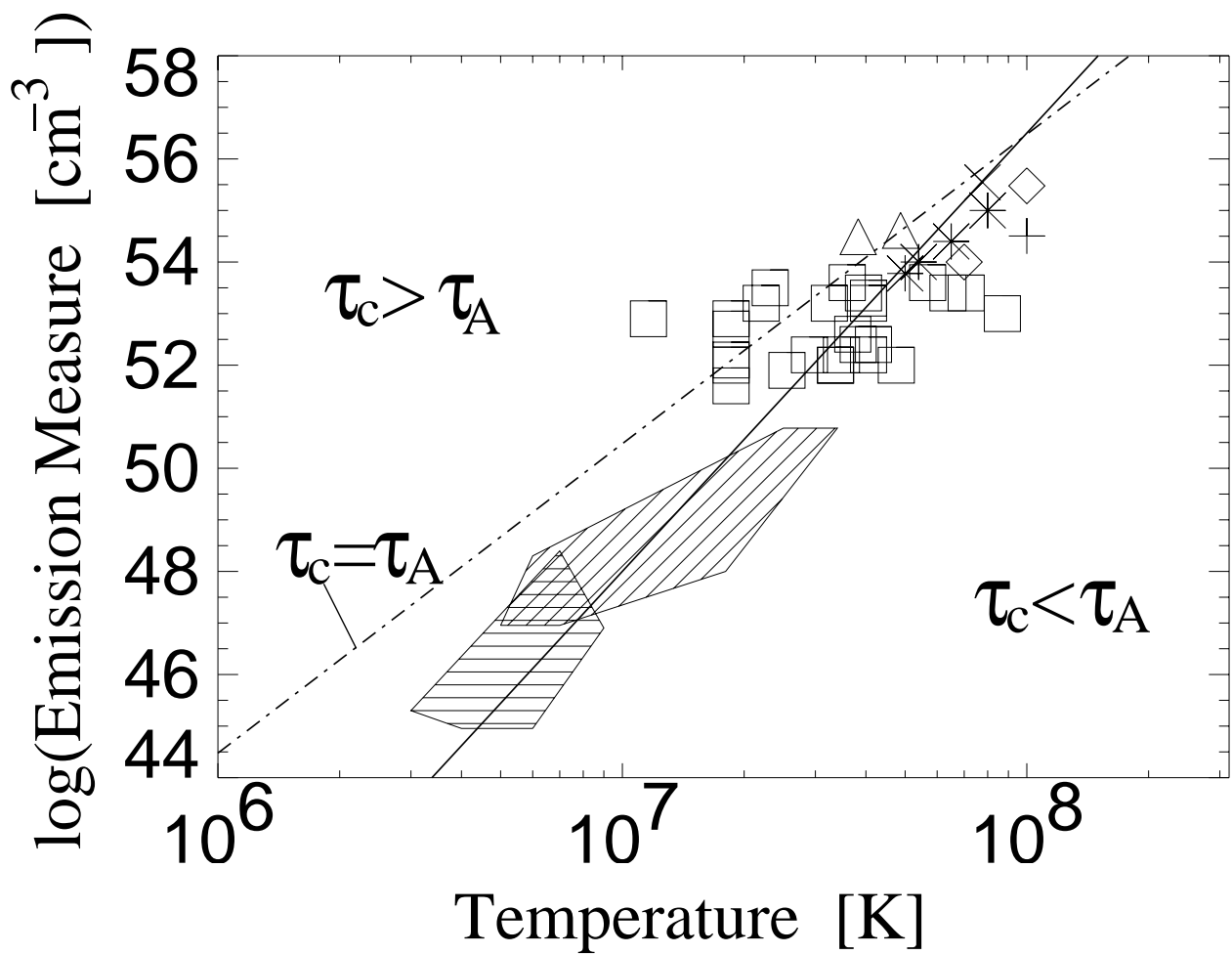

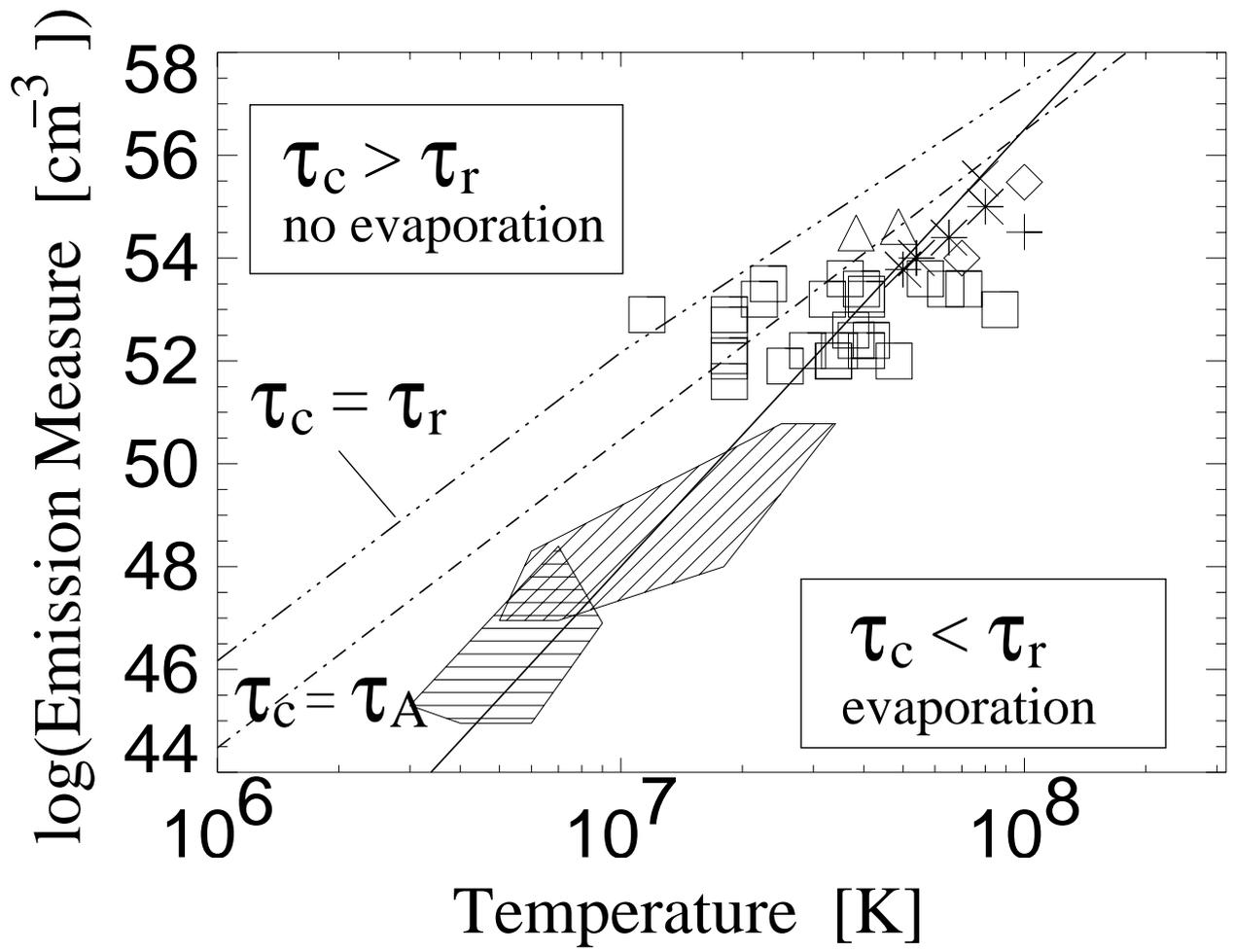